\documentclass[12pt,a4paper]{article}
\usepackage{graphicx}
\begin{document}

\title{
Recursive function templates as a solution of linear algebra expressions in C++
}
\author{{\it Volodymyr Myrnyy}\\
\normalsize{Brandenburg University of Technology, Cottbus, Germany}\\
{\normalsize\tt myrnyy@math.tu-cottbus.de}
}
\date{}
\maketitle

\abstract{
The article deals with a kind of recursive function templates in C++, where
the recursion is realized corresponding template parameters to achieve better
computational performance. Some specialization of these template functions
ends the recursion and can be implemented using optimized hardware dependent
or independent routines. The method is applied in addition to the known
expression templates technique to solve linear algebra expressions with
the help of the BLAS library. The whole implementation produces a new library,
which keeps object-oriented benefits and has a higher computational speed
represented in the tests.
}

\section*{Introduction}
The C++ programming language has a powerful template facility that enables
the development of flexible software
without incurring a large abstraction penalty \cite{Barton},\cite{tl1}.
The main goal is the resolving of all templates of a program during the
compilation time. In this way C++ language can be meant as a two-level
language \cite{Veld2}. A function template takes both template parameters
(solved in the compilation time) and function arguments (work
dynamically in the program code).

A naive implementation of linear algebra operations in C++ using the known
object-oriented features, such as providing of classes and operator
overloading, yields an inefficient code. The main reason for that is
generation of temporary objects by each overloaded operator
(\ref{fig_vmoper}). This problem is usually solved by {\it expression
templates} technique \cite{Veld1},\cite{Yang}, which is implemented
in the known optimized C++ libraries for linear algebra, e.g. {\tt valarray}
standard library (Linux RedHat 7.2), Blitz++ etc.
Their tests and descriptions \cite{blitz} prove
that computational performance of such libraries can be equal to ones
written in FORTRAN
without any object-oriented limitations.

But there are some processor and cash oriented
implementations, which have better performance. The best example of
such linear algebra library is {\bf B}asic {\bf L}inear {\bf A}lgebra
{\bf S}ubprograms \cite{blas}. They exist for most of the hardware
platforms with the same
interface, although specifically implemented (in FORTRAN or Assembler)
for some types of commonly used processors.

\begin{equation}\label{fig_vmoper}
\begin{array}{lc}
\mathbf{z}= & \underbrace{c\,\mathbf{x}} + \:\mathbf{y} \\
\mathbf{z}= & \underbrace{\mathbf{t}_1 + \mathbf{y}} \\
\mathbf{z}= & \: \mathbf{t}_2
\end{array}
\end{equation}

Therefore the combination of an expression templates abstraction with
the incorporation of BLAS in a specialization can produce a better performance
than the best pure C++ libraries.

\section*{Expression templates}
We consider a simple example of a vector expression (\ref{eq_1}) to see,
how the
expression templates are used to collect the arguments and the operators
of the vector expression.
\begin{equation}\label{eq_1}
X=A+B+C.
\end{equation}

As assumed in C++, the right hand side of the equation (\ref{eq_1}) is
resolved from left to right before the assign operator is applied. So,
we have the series of operators:
\begin{enumerate}
\item $X=A+B+C$. The first operator '+' is applied.
\item $X=BinClos<A,B,+>+C$. The second operator '+' is applied.
\item $X=BinClos<BinClos<A,B,+>,C,+>$. The assign operator '=' is applied.
\end{enumerate}
We have just introduced a new symbolic notation {\it BinClos} for a binary
operation. It looks intentionally like a template class, where
$A$, $B$ and '+' must be data types as template parameters. The basic
trick of the approach is to substitute a template class
as a template parameter into itself and to build parse trees using
operator overloading \cite{Veld1}.

The C++ class for description of binary expression closure is defined using
three template parameters: the right and the left hand side and the operation.
But it has also to save references (not the values) to the arguments of the
binary expression:
\begin{verbatim}
template<class Left, class Right, class Oper>
struct BinClos {
   const Left&  arg1;
   const Right& arg2;
   BinClos(const Left& a, const Right& b):arg1(a),arg2(b) {}  //constructor
};
\end{verbatim}
Now we have to declare the class for a vector, which can be implemented later.
Further, the addition operator should be described as a data type.
The simple structure
includes only the function {\tt apply} to realize the addition:
\begin{verbatim}
class Vector;     // some vector class
struct add {      // encapsulates the '+' operation
   static double apply(double a, double b) {
      return a+b;
   }
};
\end{verbatim}
For the whole minimal implementation we need a C++ operator, that yields the
structure {\tt BinClos}. There are some possibilities to
define this operator. The following example represents an addition of any
object of type {\tt Left} with a vector:
\begin{verbatim}
template<class Left>
BinClos<Left,Vector,add> operator + (Left& a, Vector& b) {
   return BinClos<Left,Vector,add>(a,b);
}
\end{verbatim}
So, the right hand side of the equation (\ref{eq_1}) is gathered in the
compile time to the single template structure:
\begin{equation}\label{eq_11}
\mbox{\tt BinClos<BinClos<Vector,Vector,add>,Vector,add>}
\end{equation}

The next step in the solution of (\ref{eq_1}) is to assign
the last complicated template structure (\ref{eq_11}) to the
resulting vector $X$.
We consider, at first, the usual approach that exists in the optimized C++
libraries, such as {\tt valarray} and Blitz++ \cite{Veld1}. It uses the
operator overloading to assign the whole expression in only one loop
per component (\ref{eq_2}).
\begin{equation}\label{eq_2}
X_i=A_i+B_i+C_i.
\end{equation}
The structure {\tt BinClos} is supplemented with an {\tt operator[]}, that
adds $i$-th components of two data members {\tt arg1} and {\tt arg2} of
the structure. If the {\tt arg1} (or {\tt arg2}) is a simple vector, than the
{\tt operator[]} is called in the class {\tt Vector}, else the same operator
is called recursively in the structure {\tt BinClos}. That process is started
by {\tt operator=}, which is represented by the single loop and calls
{\tt operator[]} due to the {\tt expr[i]}:
\begin{verbatim}
template<class Left, class Right, class Oper>
struct BinClos {
   const Left&  arg1;
   const Right& arg2;
   BinClos(const Left& a, const Right& b):arg1(a),arg2(b) {}
   double operator[](int i) {
      return Oper::apply(arg1[i],arg2[i]);
   }
};
class Vector
   double* data;
   int     size;
public:
   ...  //definition of constructors
   template<class Left, class Right, class Oper>
   void operator=(const BinClos<Left,Right,Oper>& expr) {
      for (int i=0; i<size; ++i) data[i]=expr[i];
   }
   double operator[](int i)  { return data[i]; }
};
\end{verbatim}

As it can be seen, the solution is distributed to some overloaded operators.
Therefore, there is no possibility to substitute an external optimized
subprogram. The next approach provides another template technique to
realize the last step of the solution.

\section*{Recursive function templates}

\begin{figure}[ht]
\center
\includegraphics[width=8cm]{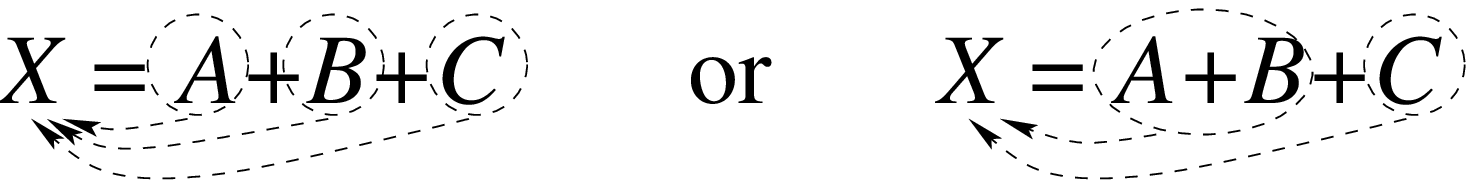}
\caption{\small Basic idea of the recursive solution}
\label{fig_idea}
\end{figure}

The basic idea of the following approach is to divide the
right hand side of (\ref{eq_1}) into simple units, which are bounded
by addition or subtraction
and than to apply the operations consequently to the vector
$X$ (fig. \ref{fig_idea}).
We have always two
types of arguments and a type of operation at the top level of
recursive built structure (\ref{eq_11}).
These two types can be applied
to the $X$ with the type of the operation. If an argument is complicated,
than the process continues in the same way recursively, else we get a simple
addition or subtraction of a vector $X$ with another vector. The last
can be done in a single function without any overloading of operators.
In the next step we have to construct the recursion using such functions.
The optimal decision in terms of maintaining the efficiency
is to apply recursion with respect to the template parameters of the functions.
In order to minimize the number of the function parameters the functions are
implemented as member functions of the structure {\tt BinClos}.

We need first to implement a C++ {\it trait} \cite{Veld1}
to represent the rule of addition.
In the terms of the set theory, we have a mapping:
\begin{equation}\label{eq_3}
\{+,+\} \rightarrow \{+\}, \quad \{+,-\} \rightarrow \{-\}, \quad
\{-,+\} \rightarrow \{-\}, \quad \{-,-\} \rightarrow \{+\}.
\end{equation}
Both addition and subtraction operations are now implemented as two
empty structures and are used as template parameters for the recursive
functions. They have only a meaning of different data types.
\begin{verbatim}
struct add {};
struct sub {};
\end{verbatim}

The trait {\tt add\_rule} receives two such empty structures as template
parameters and produces the result of the data type {\tt oper}. Two members of
the mapping (\ref{eq_3}) are described by the next template instantiation:
\begin{verbatim}
template<class Op1, class Op2>
struct add_rule {
   typedef Op2 oper;
};
\end{verbatim}
and the other two are specialized:
\begin{verbatim}
template<>
struct add_rule<sub,add> {
   typedef sub oper;
};
template<>
struct add_rule<sub,sub> {
   typedef add oper;
};
\end{verbatim}

We have now all tools to implement recursive function templates.
These are functions {\tt Assign} and {\tt Operation} (either addition or
subtraction).
The first function {\tt Assign} assigns the first argument
{\tt arg1} due to the recursive call of itself (see below). If the argument is
simple, e.g. vector, than the same function has to be implemented
in the corresponding
class (e.g. class {\tt Vector}). The function {\tt Assign}
has one template parameter, which defines the return type.
Thus, it is generalized for any operation.

The second argument {\tt arg2} of the binary closure have to be added or
subtracted from {\tt x}. For this purpose, the function {\tt Operation} is
called. It receives also the second template parameter
{\tt LeftOp} to recognize the operation. The second function {\tt Operation}
is recursive
too and used especially for C++ like expressions, e.g. $X+=A-B$. In this
case the first argument $A$ is added (first line of {\tt Operation}
implementation) and the second is subtracted. The last subtraction is
obtained by
compiler from two symbols $\{+,-\}$ as data types with the help of the
addition rule trait.

The function recursion must be finished. It means that two described
template functions must be specialized as the member
functions in each of the
class (e.g. class {\tt Vector}) that occurs in expressions resolved
by this method.

\begin{verbatim}
template<class Left, class Right, class Oper>
struct BinClos {
   const Left&  arg1;
   const Right& arg2;
   BinClos(const Left& a, const Right& b):arg1(a),arg2(b) {}

   template<class Ret>
   void Assign(Ret x) {
      arg1.Assign<Ret>(x);
      arg2.Operation<Oper,Ret>(x);
   }
   template<class LeftOp, class Ret>
   void Operation(Ret x) {
      arg1.Operation<LeftOp,Ret>(x);
      arg2.Operation<typename add_rule<LeftOp,Oper>::oper,Ret>(x);
   }
};
\end{verbatim}

The specialization in the class {\tt Vector} is very simple using the
BLAS library. The recursion process is initiated also in the class {\tt Vector}
from an overloaded operator, e.g. {\tt operator=}, by corresponding
call of the template function (e.g. {\tt Assign}):

\begin{verbatim}
class Vector
   double* data;
   int     size;
public:
   ...  //definition of constructors
   template<class Left, class Right, class Oper>
   void operator=(const BinClos<Left,Right,Oper>& expr) {
      expr.Assign(data);
   }
   template<class LeftOp, class Ret>
   void Operation(Ret x) {
      cblas_daxpy(size,1.,data,1,x,1);   //sample specialization using CBLAS
   }
};
\end{verbatim}

The provided template functions can be specialized in the structure
{\tt BinClos} for some often used short algebraic expressions. It also
leads to the increase in computational efficiency. For instance:
\begin{verbatim}
template<> template<> inline void
BinClos<Vector,Vector,add>::Assign<double*>(double* x) {        // X=A+B
   . . .
}
template<> template<> inline void
BinClos<Vector,double,mul>::Operation<add,double*>(double* x) { // X+=c*A
   . . .
}
\end{verbatim}
Moreover, multiplication operations, such as vector-constant and vector-matrix
multiplications,
must necessarily be specialized, because they cannot be partially applied
to the vector $X$.

In a whole library for linear algebra we need to implement besides the binary
expression closure in the same
way a unary expression closure and a unary and a binary function closure.
It is also important to consider that the described method of recursive
functions does not work
if any unspecialized expression in
some mathematical function is substituted, for instance:
\begin{equation}
X=sin(A+B+C), \quad \mbox{ where } X,A,B,C \mbox{ are vectors.}
\end{equation}
The acceptable solution allows the library to yield a temporary vector $T$.
This vector receives the expression value
in the mathematical function ($T=A+B+C$) and than is plugged into
the function ($X=sin(T)$).
The copying of a complicated argument result to the temporary vector does not
reduce the computational performance significantly. This can be evidently
proved by the following performance tests.

\section*{Performance tests}

\begin{figure}[ht]
%\center
\includegraphics[width=8cm]{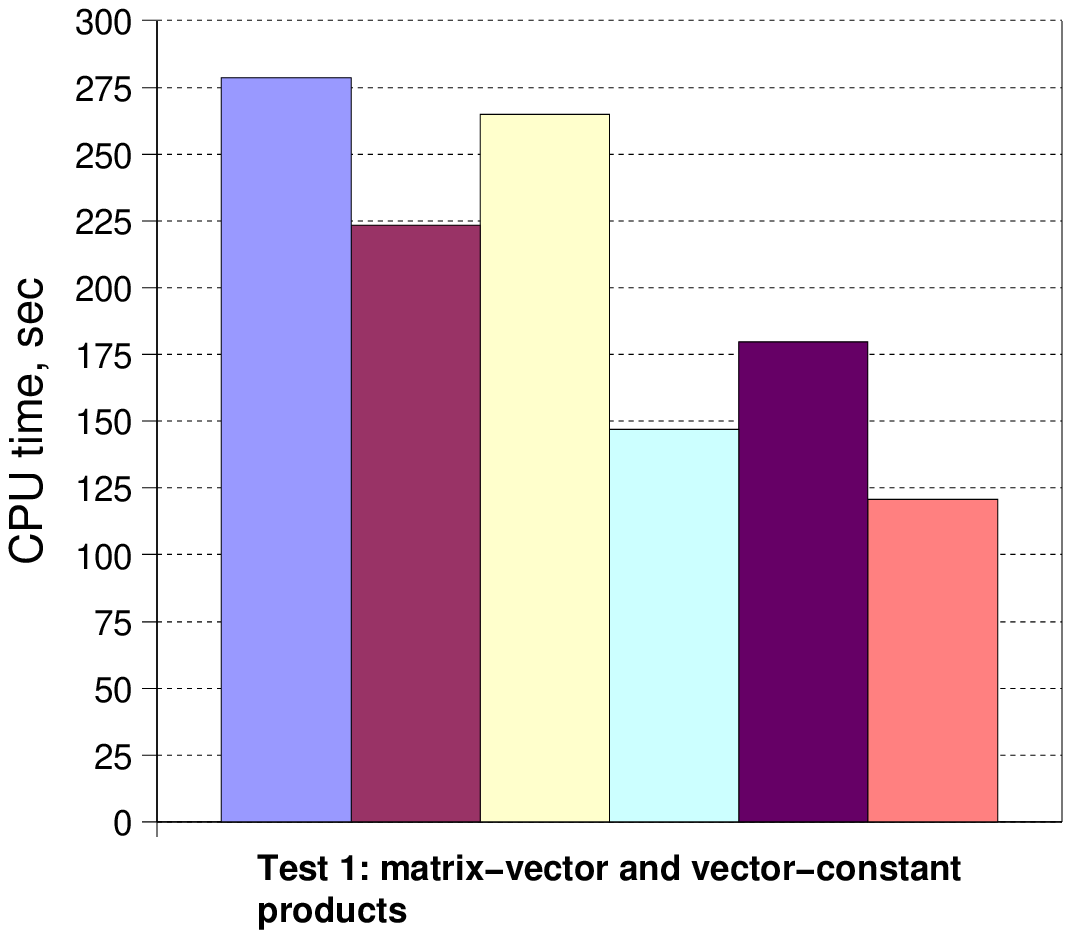}
\hspace{0.5cm}
\includegraphics[width=8cm]{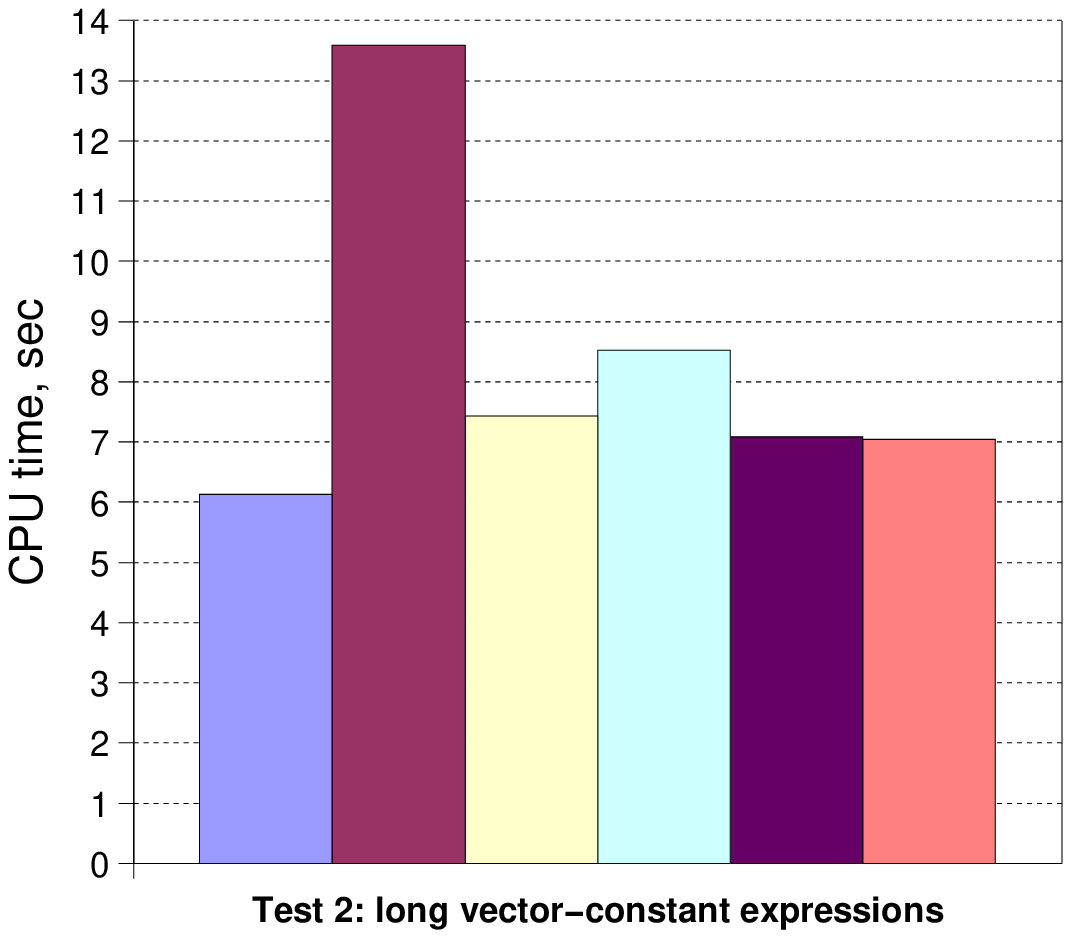}
\vspace{0.3cm}

\includegraphics[width=8cm]{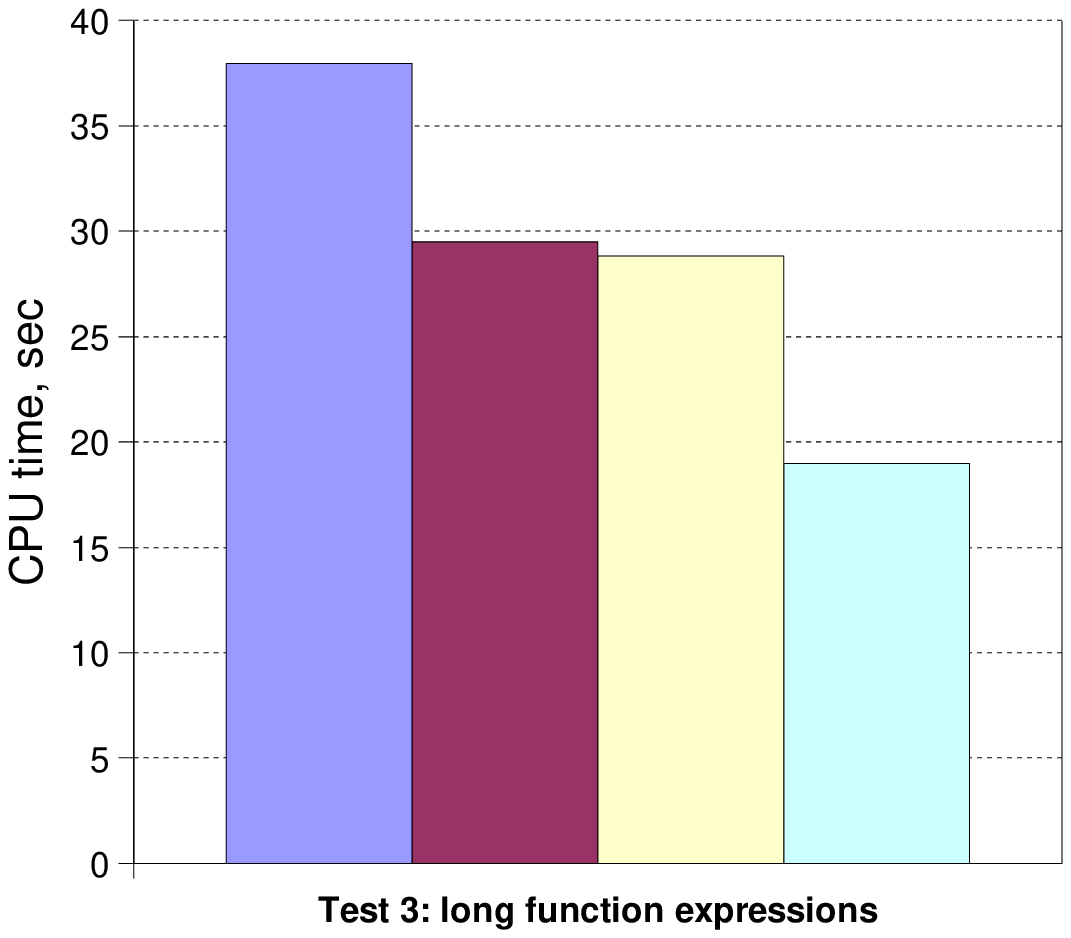}
\hspace{0.7cm}
\includegraphics[width=7cm]{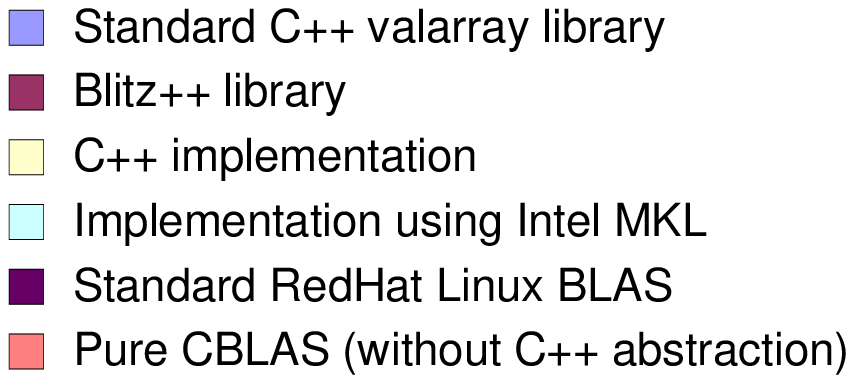}
\caption{\small Performance tests on the Intel PIII-800 platform}
\label{fig_p3}
\end{figure}

\begin{figure}[ht]
%\center
\includegraphics[width=8cm]{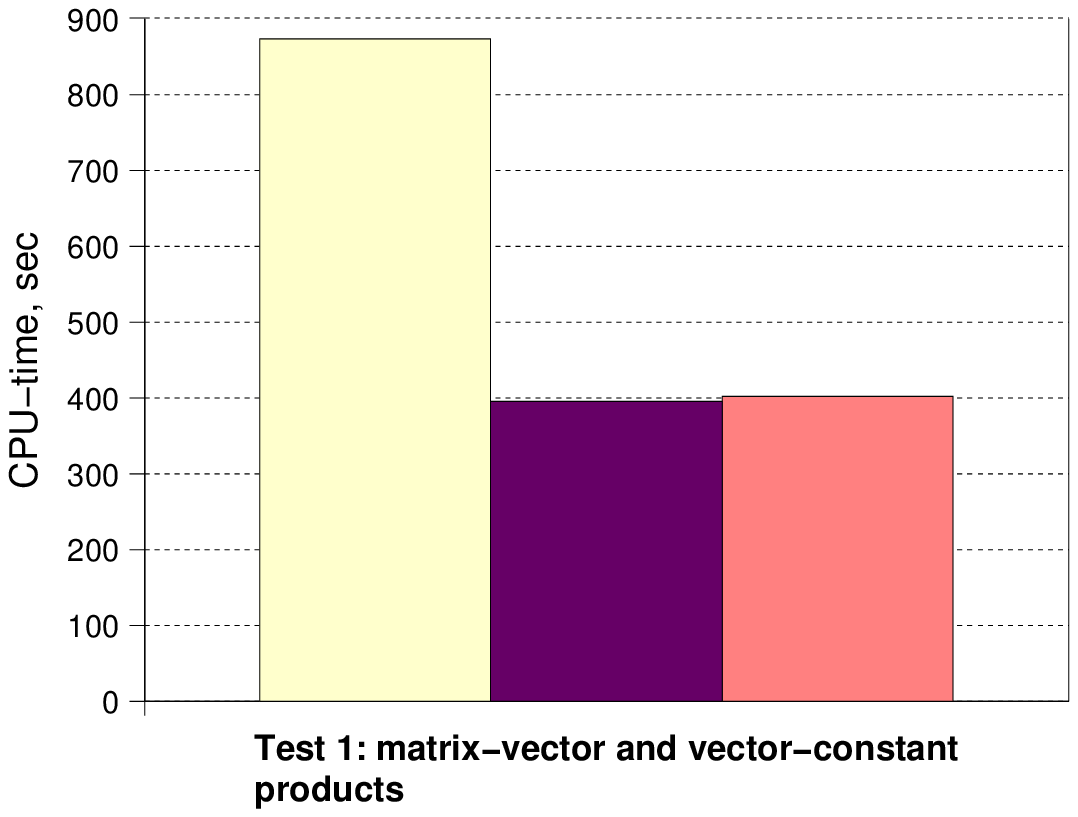}
\hspace{0.5cm}
\includegraphics[width=8cm]{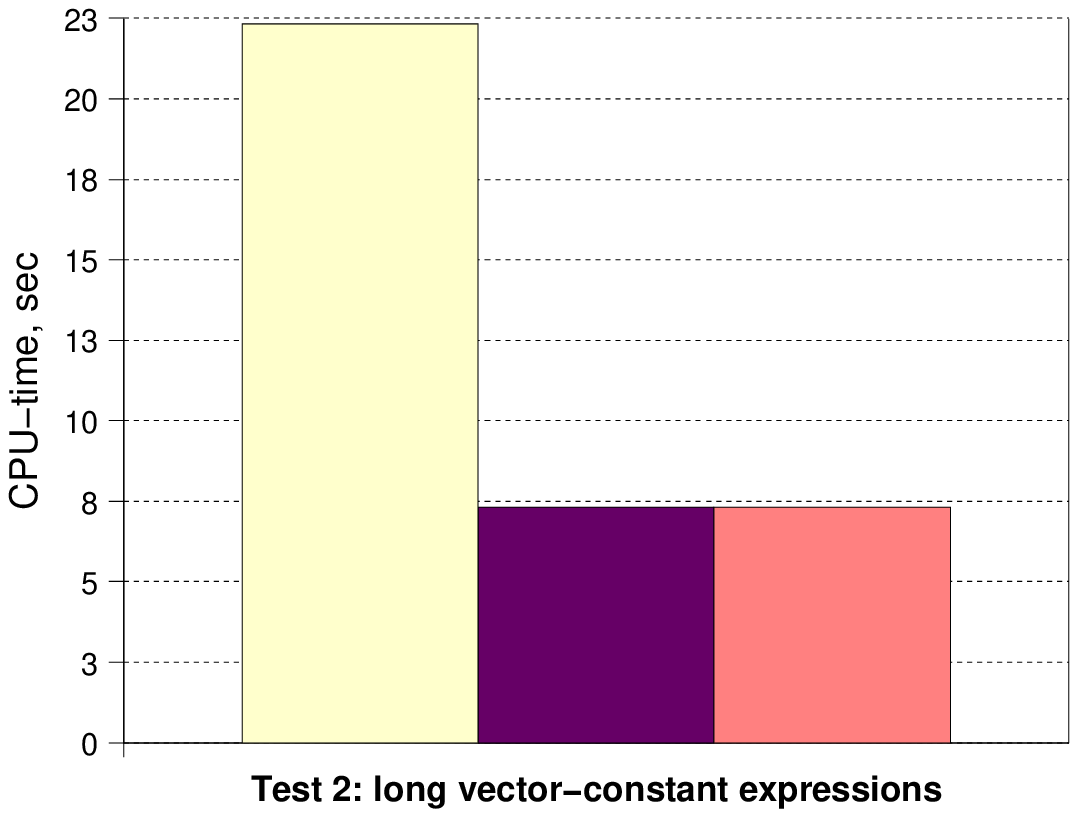}
\vspace{0.3cm}

\hspace{2.5cm}
\includegraphics[width=12cm]{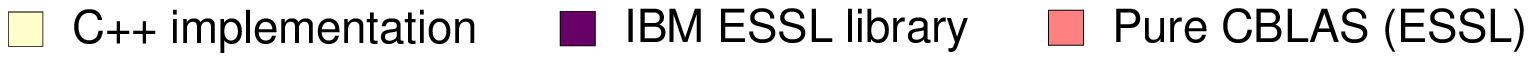}
\caption{\small Performance tests on the IBM M80 with Power3 500MHz processors}
\label{fig_m80}
\end{figure}

Three performance tests were developed to verify the functionality
of the library and to find its weak points.
\begin{itemize}
\item Test 1: Short expressions.
A vector-matrix product and a vector-constant product are repeated 1000
times. That combination is often used in many mathematical and engineering
computations.\\
$x=Ay, \qquad y=y+cx, \quad c=const$
\item Test 2: Long expressions.
The sum of 7 vector-constant products are calculated 100000 times. The test
can show how slower is the evaluation using some loops (with the help of BLAS)
and a single loop (due to overloaded {\tt operator[]}).\\
$y=y+c_1u_1+...+c_7u_7, \qquad y=c_8y, \quad c_i=const$
\item Test 3: Long function expressions.
The sum of 3 mathematical functions are calculated 50000 times. It has the same
aim as the second test. The temporary vectors in the computation of
mathematical functions and their penalty are tested.\\
$y=y+log(u_1)-cos(c_2u_2+u_3)+sin(c_4u_4+c_5u_5-u_6), \qquad y=c_6y,
\quad c_i=const$
\end{itemize}
The tests were performed on two completely different hardware platforms,
that have their own specific optimized BLAS version:
\begin{itemize}
\item Intel Pentium III 800MHz with Linux RedHat 7.3. \\
Compiler: GNU C++, gcc 2.96 version.
BLAS: Intel Math Kernel Library. \\
The first three bars of the test diagram (fig. \ref{fig_p3}) show the CPU time
of the pure C++
implementation. It proves that the new implementation (third bar) has
a performance
similar to optimized libraries {\tt valarray} and Blitz++.
The implementation using BLAS is quicker in the most cases
(test 1 and test 3),
especially using hardware specific BLAS version Intel MKL.
The last two
bars in the tests 1 and 2 estimate the abstraction penalty, that is
acceptable.
\item IBM M80 enterprise server having 8 Power3 500MHz processors and IBM AIX
operating system. (Each test uses one processor).\\
Compiler: GNU C++, gcc 2.95 version.
BLAS: IBM ESSL library.\\
The both tests (fig. \ref{fig_m80}) do not show any abstraction penalty of the
new library.
The optimized BLAS (ESSL) library requires 2-3 times less computational time
compared to the pure C++ implementation to the same task.
\end{itemize}

\clearpage

The same compiler with following optimization options was used on the both
hardware platforms:
\begin{verbatim}
-O7 -ffast-math -funroll-loops -fomit-frame-pointer -fexpensive-optimizations
\end{verbatim}

\section*{Conclusions}

The new library for the linear algebra was developed, which uses the
expression templates technique and the optimized BLAS to achieve higher
computational performance than the known C++ libraries. According to the BLAS
specification the library provides vectors and matrices of only single and
double precision
types. Any other types are not allowed, although, the
algebraic expressions resolved by the library are not limited and
implemented traditionally for both general and sparse vectors and matrices.

The main trick of the implementation is the recursion of function templates
realized by factitious empty classes {\tt add} and {\tt sub}.
%There are no any object of their type.
They have the aim to separate definition
of structure {\tt BinClos} and member function templates for addition and
subtraction operations in the compilation time. Some old compilers do not
allow to write function template if one of the template parameters is not
a data type of an argument. For some C++ developer it can seem to be a mistake.
But it works efficiently in the described implementation, since the
template parameters are resolved in the compilation time without loss
of the code performance. This code can be produced by some commonly
used compilers:
\begin{itemize}
\item GNU C++, gcc 2.95x, gcc 2.96x, gcc 3.x versions
\item Intel C++ compiler, version 5.0
\item KAI C++ compiler
\end{itemize}

\end{document}